\documentclass[%
 aip,
 amsmath,amssymb,
 reprint,%
]{revtex4-1}

\usepackage{graphicx}
\usepackage{dcolumn}
\usepackage{bm}

\usepackage[utf8]{inputenc}
\usepackage[T1]{fontenc}
\usepackage{mathptmx}

\usepackage{xcolor}
\usepackage{braket}

\usepackage{color}

\newcommand{\dd}{\text{d}}
\newcommand{\p}{\partial}
\newcommand{\grad}{\vec{\nabla}}

\newcommand{\MW}{\emph{MetalWalls}}

\newcommand{\dt}{\delta t}
\newcommand{\hdt}{\frac{\delta t}{2}}
\newcommand{\mdt}{\frac{\delta t}{m_i}}

\newcommand{\dtsq}{\delta t^2}

\renewcommand{\vec}{\bm}
\renewcommand{\rho}{\varrho}

\DeclareMathOperator{\erf}{erf}

\begin{document}

\preprint{AIP/123-QED}

\title{Mass-zero constrained molecular dynamics for electrode charges in simulations of electrochemical systems}

\author{A. Coretti}
 \affiliation{Department of Mathematical Sciences, Politecnico di Torino, I-10129 Torino, Italy}
 \affiliation{Centre Europ\'een de Calcul Atomique et Mol\'eculaire (CECAM), Ecole Polytechnique F\'ed\'erale de Lausanne, 1015 Lausanne, Switzerland}
 \author{L. Scalfi}%
\affiliation{ 
Sorbonne Universit\'e, CNRS, Physicochimie des \'Electrolytes et Nanosyst\`emes Interfaciaux, F-75005 Paris, France
}%
\author{C. Bacon}%
\affiliation{ 
Sorbonne Universit\'e, CNRS, Physicochimie des \'Electrolytes et Nanosyst\`emes Interfaciaux, F-75005 Paris, France
}%
\author{B. Rotenberg}%
\affiliation{ 
Sorbonne Universit\'e, CNRS, Physicochimie des \'Electrolytes et Nanosyst\`emes Interfaciaux, F-75005 Paris, France
}%
\affiliation{R\'eseau sur le Stockage Electrochimique de l'Energie (RS2E), FR CNRS 3459, 80039 Amiens Cedex, France}
\author{R. Vuilleumier}%
\affiliation{ 
PASTEUR, D\'epartement de chimie, \'Ecole normale sup\'erieure, PSL Univesity, Sorbonne Universit\'e. CNRS, 75005 Paris, France
}%
\author{G. Ciccotti}%
\affiliation{ 
Institute for Applied Computing ``Mauro Picone'' (IAC), CNR Via dei Taurini 19, 00185 Rome, Italy
}%
\affiliation{Universit\`a di Roma La Sapienza, Ple. A. Moro 5, 00185 Roma, Italy}
\affiliation{School of Physics, University College of Dublin UCD-Belfield, Dublin 4, Ireland}
\author{M. Salanne}
\email{mathieu.salanne@sorbonne-universite.fr}
\affiliation{ 
Sorbonne Universit\'e, CNRS, Physicochimie des \'Electrolytes et Nanosyst\`emes Interfaciaux, F-75005 Paris, France
}%
\affiliation{R\'eseau sur le Stockage Electrochimique de l'Energie (RS2E), FR CNRS 3459, 80039 Amiens Cedex, France}
\affiliation{Universit\'e Paris-Saclay, UVSQ, CNRS, CEA, Maison de la Simulation, 91191, Gif-sur-Yvette, France}
\author{S. Bonella}
 \email{sara.bonella@epfl.ch}
 \affiliation{Centre Europ\'een de Calcul Atomique et Mol\'eculaire (CECAM), Ecole Polytechnique F\'ed\'erale de Lausanne, 1015 Lausanne, Switzerland}

\date{\today}

\begin{abstract}
Classical molecular dynamics simulations have recently become a standard tool for the study of electrochemical systems. State-of-the-art approaches represent the electrodes as perfect conductors, modelling their responses to the charge distribution of electrolytes via the so-called fluctuating charge model. These fluctuating charges are additional degrees of freedom that, in a Born-Oppenheimer spirit, adapt instantaneously to changes in the environment to keep each electrode at a constant potential. Here we show that this model can be treated in the framework of constrained molecular dynamics, leading to a symplectic and time-reversible algorithm for the evolution of all the degrees of freedom of the system. The computational cost and the accuracy of the new method are similar to current alternative implementations of the model. The advantage lies in the accuracy and long term stability guaranteed by the formal properties of the algorithm and in the possibility to systematically introduce additional kinematic conditions of arbitrary number and form. We illustrate the performance of the constrained dynamics approach by enforcing the electroneutrality of the electrodes in a simple capacitor consisting of two graphite electrodes separated by a slab of liquid water.
\end{abstract}
\maketitle

\section{Introduction}

Electrochemical systems involve many complex interfaces which are difficult to characterize experimentally and to simulate accurately. Different phenomena may occur at the molecular scale, such as electron transfers, ions adsorption, chemical reactions, etc~\cite{gross2018a,fedorov2014a}. In the case of electrochemical energy storage devices, which are often operated at rather large voltages and for the largest number of cycles possible~\cite{armand2008a,simon2008a}, understanding all these processes is compulsory to design new electrode materials and/or electrolytes. From the simulation point of view, several methods can be used depending on the target properties and the scale at which the system needs to be represented. In particular, density functional theory (DFT) calculations allow to determine the thermodynamics of reactions at the close vicinity of the electrode~\cite{cheng2012a,cheng2014b,chan2015a}. Nevertheless there are many problems for which it is necessary to treat the interface at length scales going beyond the nanometer, for example in order to understand the structure of the adsorbed layer that forms and how it screens the electrode charge~\cite{fedorov2014a,salanne2016a}. In such cases, thermal fluctuations play an important role, but DFT-based molecular dynamics cannot be used because of the large system sizes that are involved (typically thousands of atoms). It is thus necessary to use a more approximate method such as classical molecular dynamics (CMD).

Indeed, CMD is commonly used to study extended systems over long time-scales. As an example, hundreds of thousands of atoms can be simulated for microseconds to study biochemical mechanisms. Yet, in most classical calculations, the molecules/materials at play are electronically insulating, so that they are  conveniently represented as collections of fixed charges over the whole simulation. This is of course not adequate to represent electrodes. More complex electrostatic schemes have been derived, among which a constant potential method was proposed by Siepmann and Sprik to account for the delocalized nature of the electronic cloud inside a metallic material and its perturbation by an external charge arising from the electrolyte~\cite{siepmann1995a}. Their method allowed to fix the potential of each electrode, therefore providing the correct framework for studying electrochemical energy storage devices. The method was later extended by Reed {\it et al.} to  the case of two-dimensional periodic systems,~\cite{reed2007a} opening the way towards systematic studies of complete electrochemical cells~\cite{merlet2012a}.

The main ingredient of the Siepmann-Sprik method is the use of fluctuating electrode charges, in the same spirit as the charge equilibration~\cite{mortier1986a,rappe1991a} scheme for molecular systems. These charges are thus treated as additional degrees of freedom in a similar way as the induced dipoles involved in polarizable force fields~\cite{madden1996a} or as the wavefunction coefficients in DFT-based molecular dynamics~\cite{vuilleumier2006a}. The same algorithms were therefore used as for these much more widely used cases. In their seminal work, Siepmann and Sprik have used the Car-Parrinello algorithm~\cite{car1985a}, which consists in introducing a fictitious mass to each additional variable and treating it as a full dynamical degree of freedom. Along the years this algorithm was progressively replaced by Born-Oppenheimer-based approaches, in which the degrees of freedom other than the atomic positions are determined either in a self-consistent way or through energy minimization techniques. Recently, a new algorithm was proposed to solve numerically the temporal evolution of polarization degrees of freedom, in the general framework of constrained molecular dynamics.~\cite{coretti2018b} The algorithm was shown to sample correctly the Born-Oppenheimer probability density and to be applicable to systems where the additional degrees of freedom must also satisfy conditions such as specific conservation or orthonormality properties, providing a viable alternative to current algorithms also for evolution based on orbital-free DFT.~\cite{bonella2020a} Here we show that this approach, named mass-zero constrained dynamics, is also very efficient for computing the electrode fluctuating charges under constant applied potential conditions, allowing further to enforce the overall electroneutrality of the system. We test the algorithm on a simple capacitor made of graphite electrodes and a pure water electrolyte. It yields instantaneous charges in perfect agreement with other techniques, and accurate properties of the system when long simulations are performed. 

\section{Theory and Algorithm}

In this section, we show how the mass-zero constrained dynamics can be used to simulate the fluctuating charge model. We begin by recalling the main features of the model, highlighting only the points more directly related to the derivation of the new approach. The extended Lagrangian associated with the mass-zero constrained dynamics and the corresponding evolution equations are then introduced. Finally, the most effective algorithm to integrate these equations for fluctuating charges, which differs from the one adopted in previous applications due to the characteristics of the model, is discussed.

The system of interest is an electrochemical cell composed of two metallic electrodes kept at constant potential and separated by an electrolyte. To set the stage, let us recall that the electric potential $V(\vec{r})$ generated by a spatial charge distribution $\rho(\vec{r})$ is given by the Poisson equation
\begin{equation}
\label{eq:poisson}
\nabla^2V(\vec{r}) = -4\pi\rho(\vec{r})
\end{equation}
complemented with appropriate boundary conditions. (Atomic units are used throughout this section.) In our case, these conditions must reflect the fact that the electrodes are conductors and therefore the potential is constant in the space they occupy. Thus
\begin{equation}\label{eq:BoundaryPot}
V(\vec{r})  = \int \dd^3 \vec{r}' \frac{\rho(\vec{r}')}{|\vec{r}-\vec{r}'|} = \Psi_{\Omega_\pm}
\end{equation}
where $\vec{r}$ is a point in the region, $\Omega_\pm$, occupied by electrode $\pm$ and $\Psi_{\Omega_\pm}$ is the known value of the potential on each electrode. An equivalent representation of the system, better suited for the introduction of the extended Lagrangian required by mass-zero constrained dynamics, is obtained by recalling that the Coulomb energy associated to the spatial charge density $\rho(\vec{r})$ can be written as
\begin{equation}
\label{eq:energy_functional}
U_c[\rho] = \frac{1}{2}\int \dd^3 \vec{r} \dd^3 \vec{r}' \frac{\rho(\vec{r}')\rho(\vec{r})}{|\vec{r}-\vec{r}'|} = \frac{1}{2}\int \dd^3 \vec{r} V(\vec{r})\rho(\vec{r}) 
\end{equation} 
The above equation implies that the electric potential can be obtained as the functional derivative of the Coulomb energy and that, in particular, the boundary condition in eq.(\ref{eq:BoundaryPot}) can also be expressed as
\begin{equation}
\label{eq:constraint_functional}
\frac{\delta U_c[\rho]}{\delta \rho(\vec{r})} = \Psi_{\Omega_\pm}
\end{equation}
for $\vec{r}\in \Omega_\pm$. A more explicit expression for the Coulomb energy of the system is obtained by specifying the charge distribution of the electrolyte and of the electrodes. In analogy with standard models of charged solutions, the electrolyte charge density is represented as a set of point charges located at the positions of its $N$ atoms. This part of the charge density will be indicated as $\rho^q_i(\vec{r}) = q_i\delta^3(\vec{r} - \vec{r}_i)$ where $q_i$ is the fixed charge of particle $i$, $\delta(x)$ is the Dirac delta function and $\vec{r}_i$ are the positions of the particles of the electrolyte. The specific feature of the fluctuating charge model, on the other hand, is to replace the continuous physical charge density of the metallic electrodes with a discrete set of $M$ Gaussian charges centered at fixed locations, $\vec{R}_\alpha$, in the $\Omega_{\pm}$ regions. 
The charge distribution assigned to the site $\vec{R}_\alpha$ is given by
\begin{equation}
\label{eq:fluctuating_density}
\rho^Q_\alpha(\vec{r}) = Q_\alpha\biggl(\frac{\eta^2}{\pi}\biggr)^{\frac{3}{2}}\exp\Bigl[-\eta^2(\vec{r} - \vec{R}_\alpha)^2\Bigr]
\end{equation}
where $\eta$ is a system-dependent parameter, fixed \textit{a priori} in the model. The $Q_\alpha$, that represent the integrated charge on each electrode site, have to be determined for each configuration of the system to enforce the conditions in eq. (\ref{eq:constraint_functional}). In particular, in a molecular dynamics simulation of the capacitor, these variables must be determined at each time step, as they respond to the evolution of the coordinates of the particles in the electrolyte along the trajectory.
Substituting the total charge density $\rho(\vec{r}) = \sum_{\alpha=1}^M\rho_\alpha^Q(\vec{r}) + \sum_{i=1}^N\rho_i^q(\vec{r})$ in eq.~\eqref{eq:energy_functional} and performing the integrals in the second equality, leads to the following explicit expression for the Coulomb energy
\begin{equation}
\begin{aligned}
\label{eq:energy_fluctuating}
U_c(\vec{r}, Q) &= \sum_{\alpha=1}^M\sum_{\beta<\alpha}\frac{Q_\alpha Q_\beta}{R_{\alpha\beta}}\erf[\sqrt{2}\eta R_{\alpha\beta}] + \sum_{\alpha=1}^N \frac{\eta}{\sqrt{2\pi}}Q_\alpha^2 \\
&+ \sum_{\alpha=1}^M\sum_{i=1}^N\frac{Q_\alpha q_i}{|\vec{R}_\alpha - \vec{r}_i|}\erf[\eta |\vec{R}_\alpha - \vec{r}_i|] + \sum_{i=1}^N\sum_{j<i}\frac{q_i q_j}{r_{ij}}
\end{aligned}
\end{equation}
where we have used the notation $\vec{r} \equiv \{\vec{r}_i \}$, $Q\equiv \{Q_{\alpha} \}$, and where $R_{\alpha\beta} = |\vec{R}_\alpha - \vec{R}_\beta|$, $r_{ij} = |\vec{r}_i - \vec{r}_j|$. Note that in this potential the $\vec{R}_{\alpha}$ are time-independent parameters, specified by the geometry of the model. Given the assumed form of the charge density on the electrodes, the condition in eq.~\eqref{eq:constraint_functional} can be reformulated in terms of the discrete ``variational'' parameters $Q_{\alpha}$ as
\begin{equation}
\label{eq:constraint_fluctuating}
\frac{\partial U_c(\vec{r}, Q)}{\partial Q_{\alpha}} = \Psi_\pm
\end{equation}
($\alpha=1,\dots,M$).
In this discrete model for the electrode charge distribution, the condition of constant potential is enforced only at sites $\vec{R}_{\alpha}$, i.e. at the center of the Gaussian charge distributions, in contrast with~\eqref{eq:constraint_functional} which holds at all positions in the metal. Previous work~\citep{reed2007a} has shown that, in spite of the approximate treatment of the charge distribution, this still enables accurate simulations of electrochemical system.

Recently, it was suggested to modify the original fluctuating charge model by forcing the $Q$ variables to respect also the so-called electroneutrality condition
\begin{equation}
\label{eq:additional_condition}
f(Q) = \sum_{\alpha=1}^MQ_\alpha = 0
\end{equation}
for all accessed configurations of the electrolyte. The condition above mimics the presence of an ideal generator connected to the electrodes and enables a closed description of the system, i.e. one that does not require to take explicitly into account interactions with a charge reservoir. As discussed in previous work\cite{scalfi2020a}, eq.~\eqref{eq:additional_condition} also ensures that relevant physical properties such as the capacitance depend only on the potential difference between the electrodes and not on their absolute potential. In simulations, this condition has the added benefit to simplify the Ewald treatment of electrostatic interactions. While electroneutrality is only enforced on average by real generators, the more stringent condition~\eqref{eq:additional_condition} has been used in simulations~\cite{limmer2013a,scalfi2020a} providing accurate results.

The dynamics of the system within the fluctuating charge model is then defined as follows. The charges $Q_{\alpha}$, mimicking the reorganization of the electronic charge density within the electrodes, adapt instantaneously to the configuration of the particles in the electrolyte. The evolution of the electrolyte's degrees of freedom is then governed by 
\begin{equation}\label{eq:StandardEvolution}
m_{i}\ddot{\vec{r}}_i = -\grad_{\vec{r}_i}U_c(\vec{r}, Q)\bigr|_{Q = \hat{Q}}
\end{equation}
where $m_i$ is the mass of the electrolyte particle $i$ and $\hat{Q} = \{\hat{Q}_1,\dots,\hat{Q}_M\}$ indicates the values of the electrode charges that satisfy eq.~\eqref{eq:constraint_fluctuating} and~\eqref{eq:additional_condition}. As mentioned in the Introduction, this dynamical model is formally identical to that of other adiabatically separated systems such as classical polarizable models or \textit{first principle} molecular dynamics. Consequently, algorithms such as Car-Parrinello dynamics or minimization schemes --- such as conjugate gradient --- typically adopted in Born-Oppenheimer propagation have been employed also to solve the evolution of the fluctuating charge model. In the following, we show how to adapt the recently proposed mass-zero constrained dynamics to this problem. In this approach an extended dynamical system is defined that leads to the same evolution equations as above for the electrolyte degrees of freedom. On the other hand, the constrained evolution of a set of auxiliary dynamical variables is used to satisfy exactly the conditions in the model. Lagrangian mechanics offers the most suitable framework to discuss the method in detail. We begin by observing that the constraints~\eqref{eq:constraint_fluctuating} can be trivially interpreted as the requirement that the function $\mathcal{W}(\vec{r}, Q)\equiv U_c(\vec{r}, Q) - \sum_{\alpha=1}^M\Psi_{\alpha}Q_{\alpha}$ (where we indicated with $\Psi_{\alpha}$ the potential at site $\vec{R}_{\alpha}$) is at a minimum with respect to $Q$. Eq.~\eqref{eq:additional_condition}, however, implies that not all variations of the $Q_{\alpha}$ are independent, and this must be accounted for when stating the minimum condition. This is conveniently done employing the formalism of Lagrange multipliers. To that end, we introduce the auxiliary function
\begin{equation}
\label{eq:extended_potential}
\mathcal{U}(\vec{r}, Q, \nu) = \mathcal{W}(\vec{r}, Q) + \nu f(Q)
\end{equation}
where $\nu$ is the Lagrange multiplier associated to the electroneutrality condition. It is easy to see, looking at the definition of $\mathcal{W}(\vec{r}, Q)$ and of $f(Q)$, that $\nu$ actually corresponds to a potential shift as already noted in previous work~\cite{scalfi2020a}. The minimum solution $\hat{Q}$ with $Q$ satisfying also eq.~\eqref{eq:additional_condition} is then given by the stationary point $\{\hat{Q}, \hat{\nu}\}$ of $\mathcal{U}(\vec{r}, Q, \nu)$. This leads to the $M+1$ conditions 
\begin{equation}\label{eq:ExtConstraints}
\begin{aligned}
\sigma_\beta(\vec{r}, Q, \nu) &= \frac{\partial \mathcal{U}(\vec{r}, Q, \nu)}{\partial Q_\beta} = 0 \quad \beta = 1,\dots,M \\
\sigma_{M+1}(Q) &= \frac{\partial \mathcal{U}(\vec{r}, Q, \nu)}{\partial \nu} = 0
\end{aligned}
\end{equation}
As usual in the Lagrange multiplier scheme, the last equation above is in fact the electroneutrality condition but now obtained as a result of an optimization problem in the space that includes the Lagrange multiplier $\nu$ as a variable. Proceeding in analogy with the derivations of the mass-zero constrained dynamics presented in ref.~\citenum{coretti2018b, bonella2020a}, we then extend the space of the dynamical degrees of freedom to include the variables $Q$ and $\nu$ and define the Lagrangian of the extended system as 
\begin{equation}
\mathcal{L} = \frac{1}{2}\sum_{i=1}^Nm_{i}\dot{r}_{i}^2 + \frac{1}{2}\sum_{\alpha=1}^{M}\mu_Q\dot{Q}_\alpha^2 +\frac{1}{2}\mu_{\nu} \dot{\nu} - \mathcal{U}(\vec{r}, Q, \nu)
\end{equation}
In the equations above, we have introduced two (arbitrary) masses, $\mu_Q$ and $\mu_{\nu}$, associated with the variables $Q$ and $\nu$, respectively. To set the stage for the homogeneous limit that we will take on these quantities (see discussion below eq.~\eqref{eq:ConstrainedSystem}), we set $\frac{\mu_Q}{\mu_v}\equiv \kappa$, with $\kappa$ a constant of appropriate physical dimensions. Conditions~\eqref{eq:ExtConstraints} are then interpreted as a set of $M+1$ holonomic constraints to obtain the evolution equations
\begin{equation}\label{eq:ConstrainedSystem}
\begin{aligned}
m_{i}\ddot{\vec{r}}_{i} &= -\grad_{\vec{r}_i}\mathcal{U}(\vec{r}, Q, \nu) + \sum_{\beta=1}^{M+1}\lambda_\beta\grad_{\vec{r}_i}\sigma_\beta \\
\mu_Q\ddot{Q}_\alpha &= - \frac{\partial \mathcal{U}(\vec{r}, Q, \nu)}{\partial Q_\alpha} + \sum_{\beta=1}^{M+1}\lambda_\beta\frac{\partial \sigma_\beta}{\partial Q_\alpha} \\
\frac{\mu_{Q}}{\kappa}\ddot{\nu}&= - \frac{\partial \mathcal{U}(\vec{r}, Q, \nu)}{\partial \nu} + \sum_{\beta=1}^{M+1}\lambda_\beta\frac{\partial \sigma_\beta}{\partial \nu}
\end{aligned}
\end{equation}
where the $\lambda_{\beta}$ are the (undetermined) Lagrange multipliers that enforce the constraints. The first derivatives of $\mathcal{U}(\vec{r}, Q, \nu)$ in the second and third line of the equations above are null due to the conditions imposed, eq.~\eqref{eq:ExtConstraints}. The equations for the $Q$ and $\nu$ variables can then be reorganized by dividing the non-vanishing terms (i.e. the constraint forces) by the masses of these degrees of freedom. In the third equation above, this means dividing by $\frac{\mu_Q}{\kappa}$. At this stage, the key step of the method is performed taking the limit $\mu_Q\rightarrow 0$ (see also ref.~\citenum{ryckaert1981a}). In order for the auxiliary variables to have a finite acceleration in this limit, the ratio $\lambda_\beta/\mu_Q$ must remain finite implying that the $\lambda_\beta$ are proportional to the masses. In the limit, the constraint forces acting on the physical degrees of freedom vanish and the dynamical system takes the form
\begin{equation}\label{eq:MaZeEquations}
\begin{aligned}
m_{i}\ddot{\vec{r}}_{i} &= -\grad_{\vec{r}_i}U_c(\vec{r}, Q) \\
\ddot{Q}_\alpha &= \sum_{\beta=1}^{M+1}\gamma_\beta\frac{\partial \sigma_\beta}{\partial Q_\alpha} \\
\ddot{\nu} &= \sum_{\beta=1}^{M+1}\gamma_\beta \kappa \frac{\partial \sigma_\beta}{\partial \nu},
\end{aligned}
\end{equation}
with $\gamma_\beta=\lim_{\mu_Q\rightarrow 0} \frac{\lambda_\beta}{\mu_Q}$ finite. In the first equation above, we used the fact that, given eq.~\eqref{eq:extended_potential} and the definition $\mathcal{W}(\vec{r}, Q)\equiv U_c(\vec{r}, Q) - \sum_{\alpha=1}^M\Psi_{\alpha}Q_{\alpha}$, we have $\grad_{\vec{r}_i}\mathcal{U}(\vec{r}, Q, \nu) = \grad_{\vec{r}_i}U_c(\vec{r}, Q)$.

Eq.~\eqref{eq:MaZeEquations} defines the mass-zero constrained dynamics for the fluctuating charge model with electroneutrality constraint. The evolution equations for the physical degrees of freedom, $\vec{r}$, do not depend directly on the constraints --- consistent with eq.~\eqref{eq:StandardEvolution}. The constrained evolution of the auxiliary variables, $\{Q,\nu\}$, however, guarantees that the stationary conditions for $\mathcal{W}(\vec{r}, Q)$ are satisfied together with the electroneutrality constraint. The zero mass limit for the auxiliary variables is rigorously enforced by this dynamical system, leading to full adiabatic separation of the physical and auxiliary motions and exact sampling of the Born-Oppenheimer probability density of the system~\cite{bonella2020a}.
The homogeneity parameter $\kappa$ introduced to take the limit of zero masses is a free parameter in the derivation of these equations of motions. To fix this parameter, one can note that it has the dimension $[\kappa]=\frac{[\nu]^2}{[Q]^2}$, that is the inverse of the square of a capacitance. The numerical value of the parameter can then be estimated from the typical capacitance for the system, e.g. the capacitance of the empty cell, $C_{\text{empty}}$. Exploratory calculations, however, showed that the numerical stability of the algorithm is insensitive to $\kappa$ in a very wide range around this value. An alternative point of view on $\kappa$ is to consider it as a scaling parameter for the additional Lagrange multiplier $\nu$ by introducing a scaled variable $\tilde{\nu}=\frac{\nu}{\sqrt{\kappa}}$. 
This new variable then has the dimension of a charge and it is natural to set $\mu_{\tilde{\nu}}=\mu_Q$. The dynamical equations resulting from this alternative perspective are strictly equivalent to Eq.~\eqref{eq:MaZeEquations}. 

We conclude this section noting that extending the set of auxiliary variables to include $\nu$ is not the only way to enforce the electroneutrality condition. For this particular case, it is in fact trivial to identify a set of generalized coordinates constraining the system to the correct hypersurface by reducing the dimensionality of the $Q$ to $M-1$ variables and setting, for example, $Q_1=-\sum^M_{\alpha=2}Q_{\alpha}$. The method proposed in this section, however, is completely general and can be applied also in cases where the definition of appropriate generalised coordinates is problematic. The only potential limitation of the method is numerical, stemming from the computational cost of enlarging the space of auxiliary variables.



\subsection*{Implementation of the mass-zero constrained algorithm}
Numerical integration of the mass-zero constrained evolution equations combines propagation of the dynamical variables with a solver for the unknown, time dependent Lagrange multipliers $\gamma\equiv\{\gamma_{\beta}\}$. In previous work, the standard implementation of the SHAKE method, employing the iterative solution of the constraints first suggested by Berendsen in~\citenum{ryckaert1977a}, was adopted. In the following we describe the specific implementation of the method in the simulation package \MW{}, a dedicated code for the simulation of electrochemical cells. To maximise efficiency, this implementation takes into account both the previous characteristics of the code and the specific properties of the fluctuating charge model, most notably the quadratic (linear) dependence of $\mathcal{U}(\vec{r},Q,\nu)$ on $Q$ ($\nu$). This dependence implies that the constraints are (at most) linear in the dynamical variables a feature that, as discussed at the end of this section, simplifies the solution of the constraints.

The basic step of the adopted algorithm is shown in eq.~\eqref{eq:unconstrained_algorithm}. It combines velocity Verlet propagation of the physical degrees of freedom with Verlet integration, incorporating the SHAKE solution of the constraints given by \eqref{eq:SHAKE_linear} (see discussion below), for the evolution of the auxiliary variables. Velocity Verlet is the integrator of choice in \MW{}. The simpler implementation of constrained evolution via Verlet propagation, added to the code to perform the calculations reported in the next section, is sufficient since the forces acting on the physical variables depend only on coordinates. Each iteration of the algorithm requires as input the positions and momenta for the physical variables at the previous timestep, $\vec{r}(t)$, $\vec{p}(t)$, and the values of the auxiliary variables at the two previous timesteps, $Q(t), \nu(t)$ and $Q(t-\delta t), \nu(t-\delta t)$ where $\delta t$ is the timestep. The knowledge of physical positions $\vec{r}(t)$ and of the additional variables $Q(t), \nu(t)$, enables to compute the forces at time $t$ as $-\nabla_{\vec{r}_i}\mathcal{U}\bigl(\vec{r}(t), Q(t), \nu (t)\bigr)=\vec{F}_i\bigl(\vec{r}(t), Q(t)\bigr)$. The Lagrange multipliers $\gamma$ are treated as free parameters not as dynamical variables, in this algorithm. Their value at $t+\dt$ is determined by the requirement that the constraints $\sigma$ at time $t+\dt$ are satisfied up to the prescribed tolerance. The notation $\gamma(t+\delta t)$ underlines this aspect. Note that, due to the form of the overall potential $\mathcal{U}(\vec{r},Q,\nu)$ and of eqns.~\eqref{eq:ExtConstraints} the derivatives of the constraints $\frac{\p \sigma_{\beta}}{\p Q_{\alpha}}(t)$, $\frac{\p \sigma_{\beta}}{\p \nu}(t)$ are in fact constant.~\footnote{We recall that the $\vec{R}_{\alpha}$ are fixed parameters in the potential. This prescription can be relaxed to include the location of the Gaussian charges to change but this extension of the model, that can be easily incorporated in the mass-zero constrained dynamics, is not considered here.} Initial conditions are chosen via standard procedures for the physical variables (see next section), while the initialization of the auxiliary variables at $t$ and $t-\delta t$ is performed using constrained conjugate gradient minimization. 
The convergence threshold for conjugate gradient minimization is set to as close as possible to zero to satisfy the constraints up (or close) to numerical precision from the beginning of the simulation.~\footnote{Note that a different initialisation scheme was proposed in ref.~\cite{coretti2018b}. This scheme is more rigorous but its implementation in \MW{} turned out to be impractical. Use of the conjugate gradient minimisation, on the other hand, is available in the code. This choice of initialisation did not cause visible problems and has negligible additional cost.}
%
\begin{widetext}
\begin{equation}
\label{eq:unconstrained_algorithm}
\begin{aligned}
&\tilde{\vec{p}}_i = \vec{p}_i(t) + \hdt \vec{F}_i\bigl(\vec{r}(t), Q(t)\bigr) \quad \text{for} \ i=1,\dots,N \\
&\vec{r}_i(t+\dt) = \vec{r}_i(t) + \mdt \tilde{\vec{p}}_i \quad \text{for} \ i=1,\dots,N \\
&\quad\left.
\begin{aligned}
&Q_\alpha(t+\dt) = 2Q_\alpha(t) - Q_\alpha(t-\dt) + \dtsq\sum^{M+1}_{\beta=1} \gamma_\beta(t+\dt) \frac{\p \sigma_\beta}{\p Q_\alpha}(t) \quad \text{for} \ \alpha=1,\dots,M \\ 
&\nu(t+\dt) = 2\nu(t) - \nu(t-\dt) + \dtsq\sum^{M+1}_{\beta=1} \gamma_\beta(t+\dt) \kappa \frac{\p \sigma_\beta}{\p \nu}(t) \\ 
&\text{COMPUTE $\{\gamma_\alpha(t+\dt)\}_{\alpha=1}^{M+1}$ SOLVING $\{\sigma_\alpha\bigl(\vec{r}(t+\dt),Q(t+\dt),\nu(t+\delta t)\bigr) = 0\}_{\alpha=1}^{M+1}$ }\\
\end{aligned}\right\}
\qquad
\begin{gathered}
\text{SHAKE} \\
\text{ALGORITHM} \\
\text{(see below)} 
\end{gathered}
\\
&\\
&\text{COMPUTE FORCES AT TIME $t+\dt$ USING $\vec{r}(t+\dt)$ AND $Q(t+\dt)$} \\
&\vec{p}_i(t+\dt) = \tilde{\vec{p}}_i + \hdt \vec{F}_i\bigl(\vec{r}(t+\dt), Q(t+\dt)\bigr) \quad \text{for} \ i=1,\dots,N\\
\end{aligned}
\end{equation}
\end{widetext}

 As mentioned above, the SHAKE algorithm used to obtain $\gamma(t+\delta t)$ exploits the specific nature of the constraints. Let us recall that SHAKE is based on a first order expansion of the constraints from which a linear system of equations for the Lagrange multiplier is obtained. The first order expansion, which in our case involves only the auxiliary variables and is then performed at fixed $\vec{r}$, is written as
\begin{equation}
\begin{aligned}\label{eq:StartSHAKE}
0 &= \sigma_\alpha\bigl(\vec{r}(t+\dt),Q(t+\dt),\nu(t+\dt)\bigr) = \\
&= \sigma_\alpha\bigl(\vec{r}(t+\dt), Q^P + \delta Q, \nu^P + \delta \nu\bigr) \\
&= \sigma_\alpha\bigl(\vec{r}(t+\dt), Q^P, \nu^P\bigr) + \sum_{\lambda=1}^{M}\frac{\p \sigma_\alpha}{\p Q_\lambda}\biggl|_{Q^P}\delta Q_\lambda + \frac{\p \sigma_\alpha}{\p \nu}\biggl|_{\nu^P}\delta \nu
\end{aligned} 
\end{equation}
for $\alpha = 1,\dots,M+1$. In the equations above, we introduced (see also SHAKE ALGORITHM block in eq.~\eqref{eq:unconstrained_algorithm}) the notation $Q_\lambda(t+\dt) = Q^P_\lambda + \delta Q_\lambda$, with $Q^P_\lambda = 2Q_\lambda(t) - Q_\lambda(t-\dt)$, $\delta Q_\lambda = \dtsq\sum^{M+1}_\beta \gamma_\beta(t+\dt) \frac{\p \sigma_\beta}{\p Q_\lambda}(t)$ and $\nu(t+\dt) = \nu^P + \delta \nu$ where $\nu^P = 2\nu(t) - \nu(t-\dt)$ and $\delta \nu = \dtsq\sum^{M+1}_\beta \kappa\gamma_\beta(t+\dt) \frac{\p \sigma_\beta}{\p \nu}(t)$. Note that, due to the linearity of the constraints, higher order terms in the expansion above are null for the fluctuating charge model. Substituting the definition of $\delta Q_\lambda$ and $\delta \nu$ in eq.~\eqref{eq:StartSHAKE} and reorganising the expression, then leads to an exact linear system for the Lagrange multipliers $\gamma$ that can be solved as
\begin{equation}
\label{eq:SHAKE_linear}
\dtsq\gamma(t+\dt) = -\sigma\bigl(\vec{r}(t+\delta t),Q^P,\nu^P\bigr)\mathbb{A}^{-1}
\end{equation}
where the elements of the matrix $\mathbb{A}$ are defined as
\begin{equation}
\label{eq:SHAKE_matr}
\mathbb{A}_{\alpha\beta} = \sum_{\lambda=1}^M\frac{\p \sigma_\alpha}{\p Q_\lambda} \frac{\p \sigma_\beta}{\p Q_\lambda} + \kappa\frac{\p \sigma_\alpha}{\p \nu} \frac{\p \sigma_\beta}{\p \nu}
\end{equation}
(Solving for the product $\dt^2\gamma$ as done in eq.~\eqref{eq:SHAKE_linear} rather than for $\gamma$ alone is known to increase the stability of the method~\cite{ciccotti1986a}.) As mentioned above, the derivatives of the constraints, and therefore the matrix $\mathbb{A}$, are constant due to the form of the constraints. The inverse matrix required in eq.~\eqref{eq:SHAKE_linear} needs therefore to be computed only once in the simulation, enabling efficient calculation of the Lagrange multipliers by direct solution of the linear system. 

A non-trivial test calculation is presented in the next section to illustrate the properties of the algorithm described above.

\section{Validation}

\begin{figure}[hbt!]
\centering
\includegraphics[width=\columnwidth]{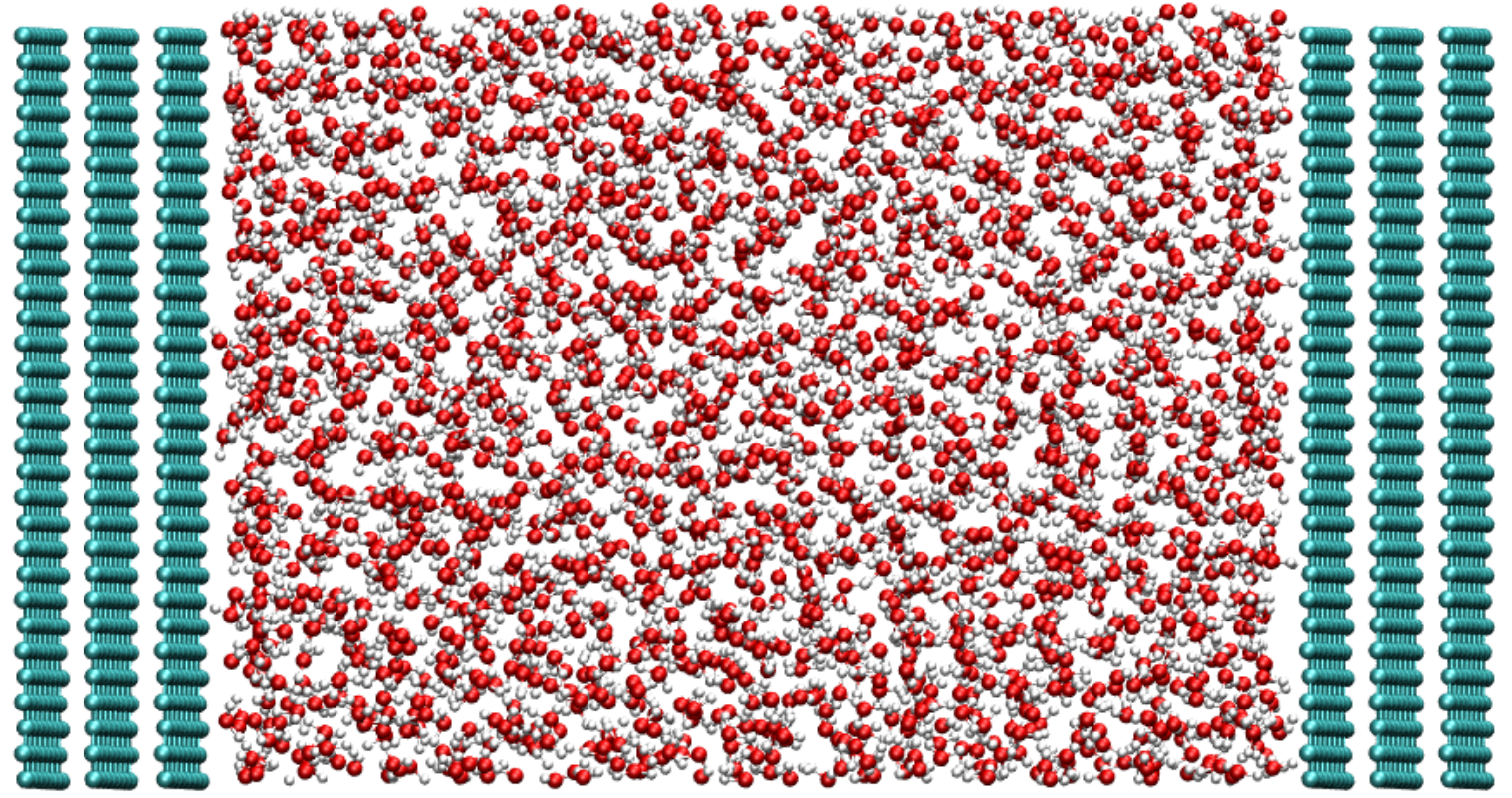}
	\caption{Representative snapshot of the simulated system (turquoise: C, red: O and white: H atoms). The $z$ direction is perpendicular to the carbon electrodes.}
\label{fig:snapshot}
\end{figure}

We simulate a system consisting of a slab of liquid water between two graphite electrodes, with the two interfaces perpendicular to the $z$ direction of the box, as shown in Figure \ref{fig:snapshot}. Despite its apparent simplicity, this system is very representative of current simulation works, which aim at characterizing the hydration of metal surfaces, both from the structural~\cite{willard2009a,limmer2013b} and the dynamical~\cite{willard2013a} points of view. The extension to more complex systems, such as dilute aqueous solutions~\cite{limmer2015b,simoncelli2018a}, ionic liquids~\cite{merlet2012a,haskins2016a,park2020a} or superconcentrated electrolytes~\cite{li2018b}, is trivial and does not require further technical development.

\subsection{Simulations}
The simulated system contains 2160 water molecules and 2880 carbon atoms in total. Each electrode is made of three graphene planes of 480 atoms each, with an interplane distance of 3.354~\AA. Two-dimensional boundary conditions were used with no periodicity in the $z$ direction. The box lengths along $x$ and $y$ were respectively fixed to $L_x$~=~34.101~\AA\ and $L_y$~=~36.915~\AA. The Lennard-Jones parameters were taken from references \citenum{berendsen1987a, werder2003a}, with the various cross-terms chosen accordingly to the Lorentz-Berthelot mixing rules. A cut-off radius of 17.05~\AA\ was used for the vdW interactions. Electrostatic interactions were computed using the two-dimensional Ewald summation technique~\cite{reed2007a,gingrich2010a}. Electrode atoms have a Gaussian charge distribution of width $\eta^{-1}$~=~0.56~\AA.

A timestep of 1~fs was used to integrate the equations of motion. The system was first equilibrated with the electrode charges set to zero at constant temperature of 298~K and atmospheric pressure. The latter condition was enforced by  allowing the positions of the electrode atoms to vary along the $z$ direction only, treating them as rigid bodies, in the presence of an external force corresponding to the target pressure. The separation between the two electrodes fluctuated around an equilibrium value of 55.11~\AA. This value was chosen to fix the positions of the graphene planes in the following simulations. The system was then equilibrated  in the NVT ensemble, with the two electrodes potential fixed to $\Psi_-=-0.5$ and $\Psi_+=0.5$~V using the conjugate gradient method. A NVE ensemble production run of 500~ps was then performed using the mass-zero constrained dynamics algorithm. The homogeneity parameter was set to $\kappa = 1$~$E_h^{2}$~$e^{-4}$ where $E_h$ is the unit of energy in Hartree atomic units and $e$ is the electron charge. The standard deviation of the total energy along the simulation is $\approx$~6$\times$10$^{-4}$~$E_h$, {\it i.e.} less than 1\% of the standard deviation of the potential energy of the system.

\subsection{Results}

In a first step, we compare the new method, which is noted as MZ in the following, to the generic Born-Oppenheimer approaches, which either employ a conjugate gradient (CG) minimization~\cite{reed2007a} or a matrix inversion~\cite{wang2014} (MI) while enforcing electroneutrality at each timestep. 

In terms of performance, a typical step using MZ only requires 1\% more CPU time than the MI method. Note that for systems with fixed positions, $\vec{R}_{\alpha}$, of the electrode's sites, these two methods are considerably faster than the CG. The conditions of the model are satisfied to a tolerance of the order of $10^{-10}$~$E_h \, e^{-1}$ with both MZ and MI. CG calculations targeting the lower, typical, tolerance of 10$^{-6}$~$E_h \, e^{-1}$ are, notably, up to a factor 5 slower. For models with fixed positions of the electrode's charges, the only numerical bottleneck for MZ and MI is the calculation of the matrix at the first time step, but this overhead is negligible for typical simulations. MZ also yields electrode charges that are in agreement with MI: we computed the charges with the two methods for 100 configurations along a trajectory and estimated the relative error $\Delta Q_\alpha$ on the electrode atom charges by computing
\begin{equation}
	\Delta Q_\alpha = \biggl|\frac{Q^{MZ}_\alpha-Q^{MI}_\alpha}{Q^{MI}_\alpha}\biggr|
\end{equation}
We obtain, on average, a relative error on the charges of $6.29\times10^{-9}$. The maximal error is $3.85\times10^{-6}$ (a single atomic charge of $-1.65120899\times10^{-9}$~$e$ instead of $-1.65121536\times10^{-9}$~$e$), which can be attributed to numerical errors.
Another important test concerns the satisfaction of the electroneutrality constraint. Along the trajectory, we obtain a maximal value for the total charge of 2.68$\times$10$^{-12}$~$e$, which is again very low and arises from the various numerical errors.

\begin{figure}[hbt!]
\centering
\includegraphics[width=\columnwidth]{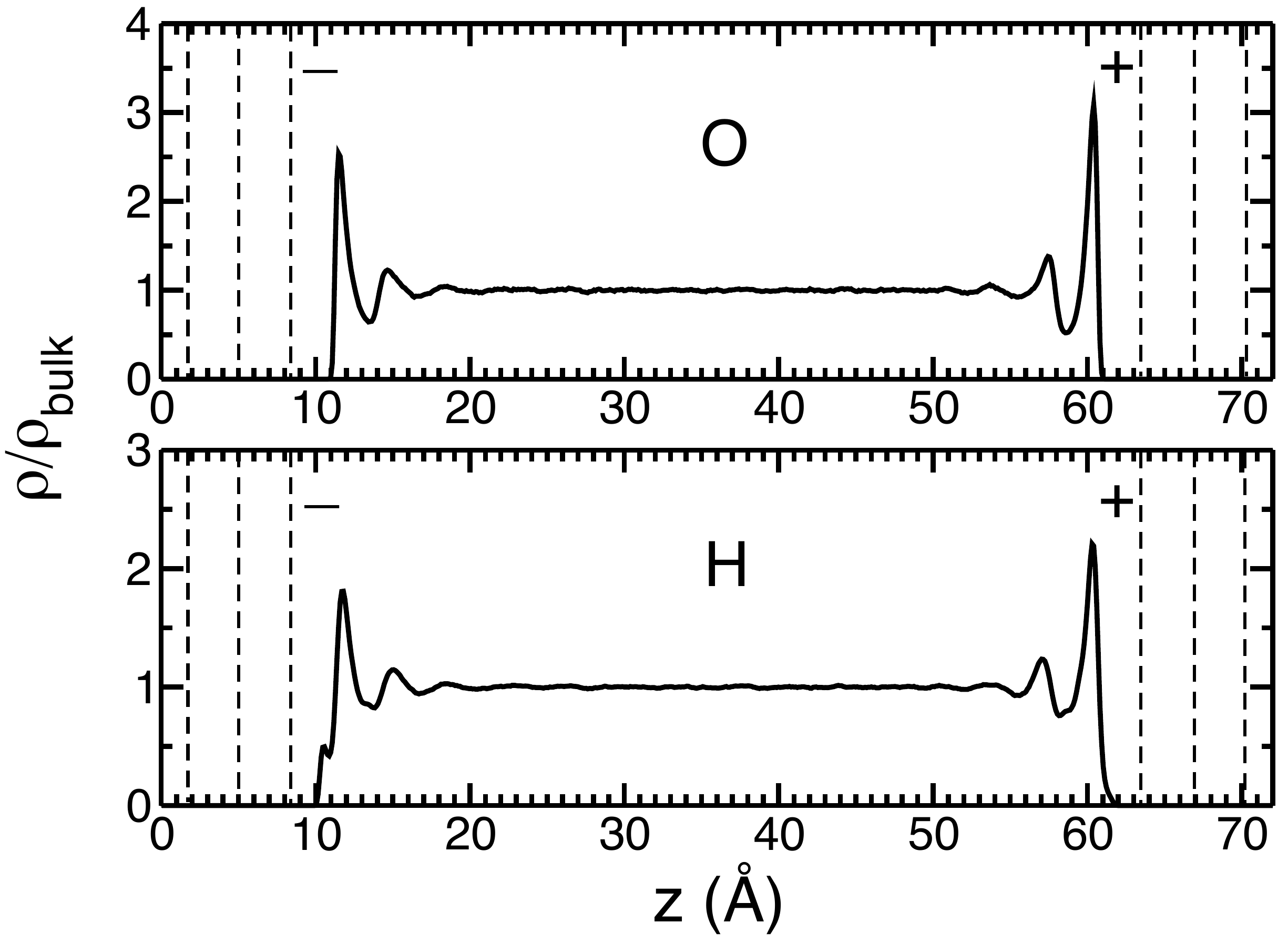}
	\caption{Normalized atomic density profiles along the $z$ direction (normal to the surface of the electrodes). The profiles are normalized by their bulk value. Top: Oxygen atoms, bottom: Hydrogen atoms. The positions of the electrode planes are shown with dashed lines.}
\label{fig:densityprofiles}
\end{figure}

Based on the excellent agreement for the calculation of the individual charges w.r.t. the MI method, we can expect the MZ to provide the correct properties of the system. To illustrate this, we first calculate the total charge on the positive electrode, which is one of the most important target properties when simulating such systems since it gives access to the interfacial capacitance. We obtain a value of 2.89~$\mu$F~cm$^{-2}$, in agreement with a previous simulation work on the same system, conducted with the CG method.~\cite{jeanmairet2019b}. 

In electrochemical devices, the structure of the liquid adsorbed at the electrode surface is difficult to characterize. Very complex experimental setups are needed, and they generally do not provide details at the molecular resolution. This structure is therefore a quantity often determined via molecular simulations. The atomic density profiles computed for our system are shown in Figure \ref{fig:densityprofiles}. These curves are also in agreement with previous studies~\cite{jeanmairet2019b} and can be qualitatively understood as follows. The presence of electrified walls leads to the formation of two layers of liquid with densities differing from the one of the bulk liquid, over a distance of approximately 10~\AA.  The density profile for the oxygen atoms shows a more intense peak close to the positive electrode (on the right hand side), while the hydrogen atoms one is characterized by the presence of a prepeak on the negative electrode side. These features are readily attributed to the polarity of water molecule. Contrarily to the case of transition metals~\cite{carrasco2012a,limmer2013b}, we do not observe the formation of particular hydrogen bonding patterns at the surface of the graphite electrode.

\begin{figure}[hbt!]
\centering
\includegraphics[width=\columnwidth]{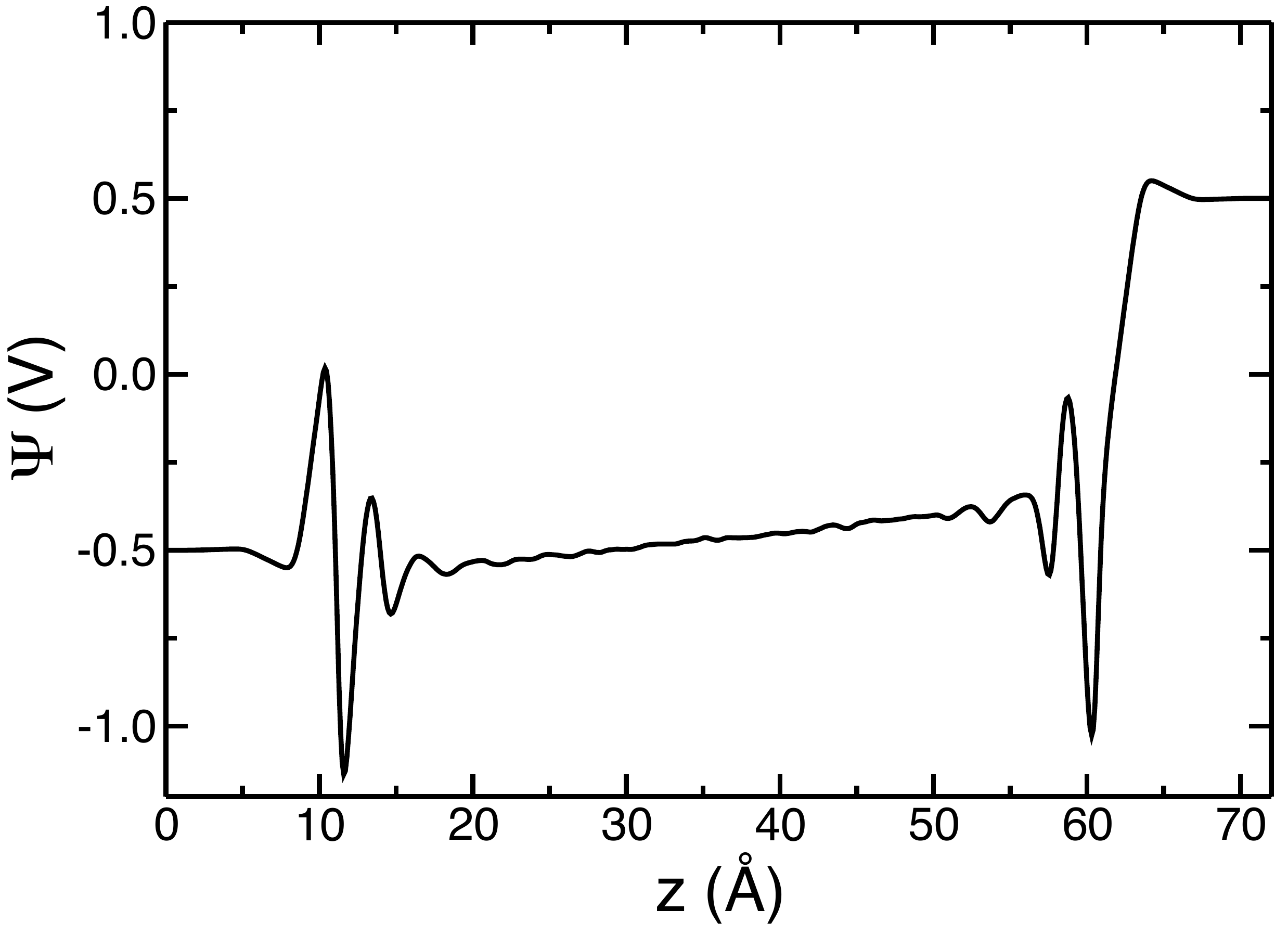}
	\caption{Poisson potential across the simulation cell.}
\label{fig:poisson_potential}
\end{figure}

The knowledge of the surface charge and of the atomic densities is not sufficient to determine the screening properties of the liquid. To better characterize the interface, it is necessary to determine the average charge density along the $z$ direction, $\braket{\rho (z)}$, in order to compute the Poisson potential across the cell, integrating twice eq.~\eqref{eq:poisson} on the $z$ axis
\begin{equation}
	\Psi(z)=\Psi(z_0)-4\pi\int_{z_0}^z {\rm d}z' \int_{-\infty}^{z'}{\rm d}z'' \braket{\rho(z'')}
\end{equation}
\noindent where  $z_0$ is a reference point. Here we choose a $z_0$ value within the left electrode, for which  $\Psi(z_0)~=~\Psi_-$~=~$-$0.5~V. As can be seen in Figure~\ref{fig:poisson_potential}, the imposed applied potential difference is recovered since the potential equals $+$0.5~V inside the right electrode. In between these two values, the Poisson potential shows several interesting features. Firstly, in the vicinity of the two electrodes, strong oscillations occur, which are due to the formation of polarized layers of water molecules. Secondly, the bulk liquid experiences a finite electric field, which differs from the applied voltage due to the screening of the adsorbed water layers.~\cite{willard2009a} By changing the applied potential, it is possible to determine further quantities such as the electric field dependence of the dielectric constant of the fluid~\cite{jeanmairet2019b}. 

\section{Conclusion}

Constant applied potential molecular dynamics simulations, which are increasingly used to study the interface between metallic electrodes and many different types of electrolytes, generally rely on a ``Born-Oppenheimer'' approximation. The models allow the electrode charges to fluctuate by enforcing the local potential to be equal to a value which is fixed for the whole electrode. In this work we have showed that it is possible to reformulate the problem, by treating these charges as dynamical variables with zero mass. The constant potential condition is recovered by using the constrained molecular dynamics formalism. This method is as efficient as the existing alternatives, but it has the advantage to allow additional conditions to be introduced in a natural way. The method also produces an algorithm which is time-reversible and symplectic for all the degrees of freedom. The new algorithm was validated by simulating a capacitor made of two graphite electrodes and a pure water simulation, with an applied potential of 1~V between the electrodes and an overall electroneutrality constraint. We have shown that the computed charges are essentially indistinguishable from those obtained via the Born-Oppenheimer approach for a given atomic configuration. We have then shown that the simulations yield structural properties of the fluid which are in perfect agreement with previous works. The results presented in this work pave the way to more general applications. In future work, in fact, the mass-zero constrained framework will be used to study systems consisting of constant potential electrodes combined with a polarizable fluid, where all the additional dynamic variables would be propagated through similar equations of motions. We also expect this method to allow fixing the charges on the two electrodes independently, thus opening the way towards comparison with more complex electrochemical experiments.

\begin{acknowledgments}

A.C. would like to acknowledge the hospitality received by PHENIX laboratories during the development of this work. This project has received funding from the European Research Council (ERC) under the European Union's Horizon 2020 research and innovation programme (grant agreement No. 771294). 
This work was supported by the French National Research Agency (Labex STORE-EX, Grant No. ANR-10-LABX-0076, and project NEPTUNE, Grant No. ANR-17-CE09-0046-02).

\end{acknowledgments}


\begin{thebibliography}{39}%
\makeatletter
\providecommand \@ifxundefined [1]{%
 \@ifx{#1\undefined}
}%
\providecommand \@ifnum [1]{%
 \ifnum #1\expandafter \@firstoftwo
 \else \expandafter \@secondoftwo
 \fi
}%
\providecommand \@ifx [1]{%
 \ifx #1\expandafter \@firstoftwo
 \else \expandafter \@secondoftwo
 \fi
}%
\providecommand \natexlab [1]{#1}%
\providecommand \enquote  [1]{``#1''}%
\providecommand \bibnamefont  [1]{#1}%
\providecommand \bibfnamefont [1]{#1}%
\providecommand \citenamefont [1]{#1}%
\providecommand \href@noop [0]{\@secondoftwo}%
\providecommand \href [0]{\begingroup \@sanitize@url \@href}%
\providecommand \@href[1]{\@@startlink{#1}\@@href}%
\providecommand \@@href[1]{\endgroup#1\@@endlink}%
\providecommand \@sanitize@url [0]{\catcode `\\12\catcode `\$12\catcode
  `\&12\catcode `\#12\catcode `\^12\catcode `\_12\catcode `\%12\relax}%
\providecommand \@@startlink[1]{}%
\providecommand \@@endlink[0]{}%
\providecommand \url  [0]{\begingroup\@sanitize@url \@url }%
\providecommand \@url [1]{\endgroup\@href {#1}{\urlprefix }}%
\providecommand \urlprefix  [0]{URL }%
\providecommand \Eprint [0]{\href }%
\providecommand \doibase [0]{http://dx.doi.org/}%
\providecommand \selectlanguage [0]{\@gobble}%
\providecommand \bibinfo  [0]{\@secondoftwo}%
\providecommand \bibfield  [0]{\@secondoftwo}%
\providecommand \translation [1]{[#1]}%
\providecommand \BibitemOpen [0]{}%
\providecommand \bibitemStop [0]{}%
\providecommand \bibitemNoStop [0]{.\EOS\space}%
\providecommand \EOS [0]{\spacefactor3000\relax}%
\providecommand \BibitemShut  [1]{\csname bibitem#1\endcsname}%
\let\auto@bib@innerbib\@empty
\bibitem [{\citenamefont {Gro\ss}(2018)}]{gross2018a}%
  \BibitemOpen
  \bibfield  {author} {\bibinfo {author} {\bibfnamefont {A.}~\bibnamefont
  {Gro\ss}},\ }\bibfield  {title} {\enquote {\bibinfo {title} {Fundamental
  challenges for modeling electrochemical energy storage systems at the atomic
  scale},}\ }\href@noop {} {\bibfield  {journal} {\bibinfo  {journal} {Top.
  Curr. Chem.}\ }\textbf {\bibinfo {volume} {376}},\ \bibinfo {pages} {17}
  (\bibinfo {year} {2018})}\BibitemShut {NoStop}%
\bibitem [{\citenamefont {Fedorov}\ and\ \citenamefont
  {Kornyshev}(2014)}]{fedorov2014a}%
  \BibitemOpen
  \bibfield  {author} {\bibinfo {author} {\bibfnamefont {M.~V.}\ \bibnamefont
  {Fedorov}}\ and\ \bibinfo {author} {\bibfnamefont {A.~A.}\ \bibnamefont
  {Kornyshev}},\ }\bibfield  {title} {\enquote {\bibinfo {title} {Ionic liquids
  at electrified interfaces},}\ }\href@noop {} {\bibfield  {journal} {\bibinfo
  {journal} {Chem. Rev.}\ }\textbf {\bibinfo {volume} {114}},\ \bibinfo {pages}
  {2978---3036} (\bibinfo {year} {2014})}\BibitemShut {NoStop}%
\bibitem [{\citenamefont {Armand}\ and\ \citenamefont
  {Tarascon}(2008)}]{armand2008a}%
  \BibitemOpen
  \bibfield  {author} {\bibinfo {author} {\bibfnamefont {M.}~\bibnamefont
  {Armand}}\ and\ \bibinfo {author} {\bibfnamefont {J.-M.}\ \bibnamefont
  {Tarascon}},\ }\bibfield  {title} {\enquote {\bibinfo {title} {Building
  better batteries},}\ }\href@noop {} {\bibfield  {journal} {\bibinfo
  {journal} {Nature}\ }\textbf {\bibinfo {volume} {451}},\ \bibinfo {pages}
  {652--657} (\bibinfo {year} {2008})}\BibitemShut {NoStop}%
\bibitem [{\citenamefont {Simon}\ and\ \citenamefont
  {Gogotsi}(2008)}]{simon2008a}%
  \BibitemOpen
  \bibfield  {author} {\bibinfo {author} {\bibfnamefont {P.}~\bibnamefont
  {Simon}}\ and\ \bibinfo {author} {\bibfnamefont {Y.}~\bibnamefont
  {Gogotsi}},\ }\bibfield  {title} {\enquote {\bibinfo {title} {{Materials for
  Electrochemical Capacitors}},}\ }\href@noop {} {\bibfield  {journal}
  {\bibinfo  {journal} {Nat. Mater.}\ }\textbf {\bibinfo {volume} {7}},\
  \bibinfo {pages} {845--854} (\bibinfo {year} {2008})}\BibitemShut {NoStop}%
\bibitem [{\citenamefont {Cheng}\ and\ \citenamefont
  {Sprik}(2012)}]{cheng2012a}%
  \BibitemOpen
  \bibfield  {author} {\bibinfo {author} {\bibfnamefont {J.}~\bibnamefont
  {Cheng}}\ and\ \bibinfo {author} {\bibfnamefont {M.}~\bibnamefont {Sprik}},\
  }\bibfield  {title} {\enquote {\bibinfo {title} {Alignment of electronic
  energy levels at electrochemical interfaces},}\ }\href@noop {} {\bibfield
  {journal} {\bibinfo  {journal} {Phys. Chem. Chem. Phys.}\ }\textbf {\bibinfo
  {volume} {14}},\ \bibinfo {pages} {11245--11267} (\bibinfo {year}
  {2012})}\BibitemShut {NoStop}%
\bibitem [{\citenamefont {Cheng}\ \emph {et~al.}(2014)\citenamefont {Cheng},
  \citenamefont {Liu}, \citenamefont {Kattirtzi}, \citenamefont
  {VandeVondele},\ and\ \citenamefont {Sprik}}]{cheng2014b}%
  \BibitemOpen
  \bibfield  {author} {\bibinfo {author} {\bibfnamefont {J.}~\bibnamefont
  {Cheng}}, \bibinfo {author} {\bibfnamefont {X.}~\bibnamefont {Liu}}, \bibinfo
  {author} {\bibfnamefont {J.~A.}\ \bibnamefont {Kattirtzi}}, \bibinfo {author}
  {\bibfnamefont {J.}~\bibnamefont {VandeVondele}}, \ and\ \bibinfo {author}
  {\bibfnamefont {M.}~\bibnamefont {Sprik}},\ }\bibfield  {title} {\enquote
  {\bibinfo {title} {Aligning electronic and protonic energy levels of
  proton-coupled electron transfer in water oxidation on aqueous {TiO}$_2$},}\
  }\href@noop {} {\bibfield  {journal} {\bibinfo  {journal} {Angew. Chem., Int.
  Ed.}\ }\textbf {\bibinfo {volume} {53}},\ \bibinfo {pages} {12046--12050}
  (\bibinfo {year} {2014})}\BibitemShut {NoStop}%
\bibitem [{\citenamefont {Chan}\ and\ \citenamefont
  {Norskov}(2015)}]{chan2015a}%
  \BibitemOpen
  \bibfield  {author} {\bibinfo {author} {\bibfnamefont {K.}~\bibnamefont
  {Chan}}\ and\ \bibinfo {author} {\bibfnamefont {J.~K.}\ \bibnamefont
  {Norskov}},\ }\bibfield  {title} {\enquote {\bibinfo {title} {Electrochemical
  barriers made simple},}\ }\href@noop {} {\bibfield  {journal} {\bibinfo
  {journal} {J. Phys. Chem. Lett.}\ }\textbf {\bibinfo {volume} {6}},\ \bibinfo
  {pages} {2663--2668} (\bibinfo {year} {2015})}\BibitemShut {NoStop}%
\bibitem [{\citenamefont {Salanne}\ \emph {et~al.}(2016)\citenamefont
  {Salanne}, \citenamefont {Rotenberg}, \citenamefont {Naoi}, \citenamefont
  {Kaneko}, \citenamefont {Taberna}, \citenamefont {Grey}, \citenamefont
  {Dunn},\ and\ \citenamefont {Simon}}]{salanne2016a}%
  \BibitemOpen
  \bibfield  {author} {\bibinfo {author} {\bibfnamefont {M.}~\bibnamefont
  {Salanne}}, \bibinfo {author} {\bibfnamefont {B.}~\bibnamefont {Rotenberg}},
  \bibinfo {author} {\bibfnamefont {K.}~\bibnamefont {Naoi}}, \bibinfo {author}
  {\bibfnamefont {K.}~\bibnamefont {Kaneko}}, \bibinfo {author} {\bibfnamefont
  {P.-L.}\ \bibnamefont {Taberna}}, \bibinfo {author} {\bibfnamefont {C.~P.}\
  \bibnamefont {Grey}}, \bibinfo {author} {\bibfnamefont {B.}~\bibnamefont
  {Dunn}}, \ and\ \bibinfo {author} {\bibfnamefont {P.}~\bibnamefont {Simon}},\
  }\bibfield  {title} {\enquote {\bibinfo {title} {{Efficient Storage
  Mechanisms for Building Better Supercapacitors}},}\ }\href@noop {} {\bibfield
   {journal} {\bibinfo  {journal} {Nat. Energy}\ }\textbf {\bibinfo {volume}
  {1}},\ \bibinfo {pages} {16070} (\bibinfo {year} {2016})}\BibitemShut
  {NoStop}%
\bibitem [{\citenamefont {Siepmann}\ and\ \citenamefont
  {Sprik}(1995)}]{siepmann1995a}%
  \BibitemOpen
  \bibfield  {author} {\bibinfo {author} {\bibfnamefont {J.~I.}\ \bibnamefont
  {Siepmann}}\ and\ \bibinfo {author} {\bibfnamefont {M.}~\bibnamefont
  {Sprik}},\ }\bibfield  {title} {\enquote {\bibinfo {title} {{Influence of
  Surface-Topology and Electrostatic Potential on Water Electrode Systems}},}\
  }\href@noop {} {\bibfield  {journal} {\bibinfo  {journal} {J. Chem. Phys.}\
  }\textbf {\bibinfo {volume} {102}},\ \bibinfo {pages} {511--524} (\bibinfo
  {year} {1995})}\BibitemShut {NoStop}%
\bibitem [{\citenamefont {Reed}, \citenamefont {Lanning},\ and\ \citenamefont
  {Madden}(2007)}]{reed2007a}%
  \BibitemOpen
  \bibfield  {author} {\bibinfo {author} {\bibfnamefont {S.~K.}\ \bibnamefont
  {Reed}}, \bibinfo {author} {\bibfnamefont {O.~J.}\ \bibnamefont {Lanning}}, \
  and\ \bibinfo {author} {\bibfnamefont {P.~A.}\ \bibnamefont {Madden}},\
  }\bibfield  {title} {\enquote {\bibinfo {title} {{Electrochemical Interface
  Between an Ionic Liquid and a Model Metallic Electrode}},}\ }\href@noop {}
  {\bibfield  {journal} {\bibinfo  {journal} {J. Chem. Phys.}\ }\textbf
  {\bibinfo {volume} {126}},\ \bibinfo {pages} {084704} (\bibinfo {year}
  {2007})}\BibitemShut {NoStop}%
\bibitem [{\citenamefont {Merlet}\ \emph {et~al.}(2012)\citenamefont {Merlet},
  \citenamefont {Rotenberg}, \citenamefont {Madden}, \citenamefont {Taberna},
  \citenamefont {Simon}, \citenamefont {Gogotsi},\ and\ \citenamefont
  {Salanne}}]{merlet2012a}%
  \BibitemOpen
  \bibfield  {author} {\bibinfo {author} {\bibfnamefont {C.}~\bibnamefont
  {Merlet}}, \bibinfo {author} {\bibfnamefont {B.}~\bibnamefont {Rotenberg}},
  \bibinfo {author} {\bibfnamefont {P.~A.}\ \bibnamefont {Madden}}, \bibinfo
  {author} {\bibfnamefont {P.-L.}\ \bibnamefont {Taberna}}, \bibinfo {author}
  {\bibfnamefont {P.}~\bibnamefont {Simon}}, \bibinfo {author} {\bibfnamefont
  {Y.}~\bibnamefont {Gogotsi}}, \ and\ \bibinfo {author} {\bibfnamefont
  {M.}~\bibnamefont {Salanne}},\ }\bibfield  {title} {\enquote {\bibinfo
  {title} {{On the Molecular Origin of Supercapacitance in Nanoporous Carbon
  Electrodes}},}\ }\href@noop {} {\bibfield  {journal} {\bibinfo  {journal}
  {Nat. Mater.}\ }\textbf {\bibinfo {volume} {11}},\ \bibinfo {pages}
  {306--310} (\bibinfo {year} {2012})}\BibitemShut {NoStop}%
\bibitem [{\citenamefont {Mortier}, \citenamefont {Ghosh},\ and\ \citenamefont
  {Shankar}(1986)}]{mortier1986a}%
  \BibitemOpen
  \bibfield  {author} {\bibinfo {author} {\bibfnamefont {W.~J.}\ \bibnamefont
  {Mortier}}, \bibinfo {author} {\bibfnamefont {S.~K.}\ \bibnamefont {Ghosh}},
  \ and\ \bibinfo {author} {\bibfnamefont {S.}~\bibnamefont {Shankar}},\
  }\bibfield  {title} {\enquote {\bibinfo {title}
  {Electronegativity-equalization method for the calculation of atomic charges
  in molecules},}\ }\href@noop {} {\bibfield  {journal} {\bibinfo  {journal}
  {J. Am. Chem. Soc.}\ }\textbf {\bibinfo {volume} {108}},\ \bibinfo {pages}
  {4315--4320} (\bibinfo {year} {1986})}\BibitemShut {NoStop}%
\bibitem [{\citenamefont {Rappe}\ and\ \citenamefont {{Goddard
  III}}(1991)}]{rappe1991a}%
  \BibitemOpen
  \bibfield  {author} {\bibinfo {author} {\bibfnamefont {A.~K.}\ \bibnamefont
  {Rappe}}\ and\ \bibinfo {author} {\bibfnamefont {W.~A.}\ \bibnamefont
  {{Goddard III}}},\ }\bibfield  {title} {\enquote {\bibinfo {title} {Charge
  equilibration for molecular dynamics simulations},}\ }\href@noop {}
  {\bibfield  {journal} {\bibinfo  {journal} {J. Phys. Chem.}\ }\textbf
  {\bibinfo {volume} {95}},\ \bibinfo {pages} {3358--3363} (\bibinfo {year}
  {1991})}\BibitemShut {NoStop}%
\bibitem [{\citenamefont {Madden}\ and\ \citenamefont
  {Wilson}(1996)}]{madden1996a}%
  \BibitemOpen
  \bibfield  {author} {\bibinfo {author} {\bibfnamefont {P.~A.}\ \bibnamefont
  {Madden}}\ and\ \bibinfo {author} {\bibfnamefont {M.}~\bibnamefont
  {Wilson}},\ }\bibfield  {title} {\enquote {\bibinfo {title} {'covalent'
  effects in 'ionic' systems},}\ }\href@noop {} {\bibfield  {journal} {\bibinfo
   {journal} {Chem. Soc. Rev.}\ }\textbf {\bibinfo {volume} {25}},\ \bibinfo
  {pages} {339--350} (\bibinfo {year} {1996})}\BibitemShut {NoStop}%
\bibitem [{\citenamefont {Vuilleumier}(2006)}]{vuilleumier2006a}%
  \BibitemOpen
  \bibfield  {author} {\bibinfo {author} {\bibfnamefont {R.}~\bibnamefont
  {Vuilleumier}},\ }\bibfield  {title} {\enquote {\bibinfo {title} {Density
  functional theory based ab initio molecular dynamics using the car-parrinello
  approach},}\ }\href@noop {} {\bibfield  {journal} {\bibinfo  {journal} {Lect.
  Notes Phys.}\ }\textbf {\bibinfo {volume} {703}},\ \bibinfo {pages}
  {223--285} (\bibinfo {year} {2006})}\BibitemShut {NoStop}%
\bibitem [{\citenamefont {Car}\ and\ \citenamefont
  {Parrinello}(1985)}]{car1985a}%
  \BibitemOpen
  \bibfield  {author} {\bibinfo {author} {\bibfnamefont {R.}~\bibnamefont
  {Car}}\ and\ \bibinfo {author} {\bibfnamefont {M.}~\bibnamefont
  {Parrinello}},\ }\bibfield  {title} {\enquote {\bibinfo {title} {Unified
  approach for molecular dynamics and density-functional theory},}\ }\href@noop
  {} {\bibfield  {journal} {\bibinfo  {journal} {Phys. Rev. Lett.}\ }\textbf
  {\bibinfo {volume} {55}},\ \bibinfo {pages} {2471--2474} (\bibinfo {year}
  {1985})}\BibitemShut {NoStop}%
\bibitem [{\citenamefont {Coretti}, \citenamefont {Bonella},\ and\
  \citenamefont {Ciccotti}(2018)}]{coretti2018b}%
  \BibitemOpen
  \bibfield  {author} {\bibinfo {author} {\bibfnamefont {A.}~\bibnamefont
  {Coretti}}, \bibinfo {author} {\bibfnamefont {S.}~\bibnamefont {Bonella}}, \
  and\ \bibinfo {author} {\bibfnamefont {G.}~\bibnamefont {Ciccotti}},\
  }\bibfield  {title} {\enquote {\bibinfo {title} {Communication: Constrained
  molecular dynamics for polarizable models},}\ }\href@noop {} {\bibfield
  {journal} {\bibinfo  {journal} {J. Chem. Phys.}\ }\textbf {\bibinfo {volume}
  {149}},\ \bibinfo {pages} {191102} (\bibinfo {year} {2018})}\BibitemShut
  {NoStop}%
\bibitem [{\citenamefont {Bonella}\ \emph {et~al.}(2020)\citenamefont
  {Bonella}, \citenamefont {Coretti}, \citenamefont {Vuilleumier},\ and\
  \citenamefont {Ciccotti}}]{bonella2020a}%
  \BibitemOpen
  \bibfield  {author} {\bibinfo {author} {\bibfnamefont {S.}~\bibnamefont
  {Bonella}}, \bibinfo {author} {\bibfnamefont {A.}~\bibnamefont {Coretti}},
  \bibinfo {author} {\bibfnamefont {R.}~\bibnamefont {Vuilleumier}}, \ and\
  \bibinfo {author} {\bibfnamefont {G.}~\bibnamefont {Ciccotti}},\ }\bibfield
  {title} {\enquote {\bibinfo {title} {Adiabatic motion and statistical
  mechanics via mass zero constrained dynamics},}\ }\href
  {http://dx.doi.org/10.1039/D0CP00163E} {\bibfield  {journal} {\bibinfo
  {journal} {Phys. Chem. Chem. Phys.}\ }\textbf {\bibinfo {volume} {in press}}
  (\bibinfo {year} {2020})}\BibitemShut {NoStop}%
\bibitem [{\citenamefont {Scalfi}\ \emph {et~al.}(2020)\citenamefont {Scalfi},
  \citenamefont {Limmer}, \citenamefont {Coretti}, \citenamefont {Bonella},
  \citenamefont {Madden}, \citenamefont {Salanne},\ and\ \citenamefont
  {Rotenberg}}]{scalfi2020a}%
  \BibitemOpen
  \bibfield  {author} {\bibinfo {author} {\bibfnamefont {L.}~\bibnamefont
  {Scalfi}}, \bibinfo {author} {\bibfnamefont {D.~T.}\ \bibnamefont {Limmer}},
  \bibinfo {author} {\bibfnamefont {A.}~\bibnamefont {Coretti}}, \bibinfo
  {author} {\bibfnamefont {S.}~\bibnamefont {Bonella}}, \bibinfo {author}
  {\bibfnamefont {P.~A.}\ \bibnamefont {Madden}}, \bibinfo {author}
  {\bibfnamefont {M.}~\bibnamefont {Salanne}}, \ and\ \bibinfo {author}
  {\bibfnamefont {B.}~\bibnamefont {Rotenberg}},\ }\bibfield  {title} {\enquote
  {\bibinfo {title} {Charge fluctuations from molecular simulations in the
  constant-potential ensemble},}\ }\href@noop {} {\bibfield  {journal}
  {\bibinfo  {journal} {Phys. Chem. Chem. Phys.}\ }\textbf {\bibinfo {volume}
  {in press}} (\bibinfo {year} {2020})}\BibitemShut {NoStop}%
\bibitem [{\citenamefont {Limmer}\ \emph
  {et~al.}(2013{\natexlab{a}})\citenamefont {Limmer}, \citenamefont {Merlet},
  \citenamefont {Salanne}, \citenamefont {Chandler}, \citenamefont {Madden},
  \citenamefont {{van Roij}},\ and\ \citenamefont {Rotenberg}}]{limmer2013a}%
  \BibitemOpen
  \bibfield  {author} {\bibinfo {author} {\bibfnamefont {D.~T.}\ \bibnamefont
  {Limmer}}, \bibinfo {author} {\bibfnamefont {C.}~\bibnamefont {Merlet}},
  \bibinfo {author} {\bibfnamefont {M.}~\bibnamefont {Salanne}}, \bibinfo
  {author} {\bibfnamefont {D.}~\bibnamefont {Chandler}}, \bibinfo {author}
  {\bibfnamefont {P.~A.}\ \bibnamefont {Madden}}, \bibinfo {author}
  {\bibfnamefont {R.}~\bibnamefont {{van Roij}}}, \ and\ \bibinfo {author}
  {\bibfnamefont {B.}~\bibnamefont {Rotenberg}},\ }\bibfield  {title} {\enquote
  {\bibinfo {title} {Charge fluctuations in nanoscale capacitors},}\
  }\href@noop {} {\bibfield  {journal} {\bibinfo  {journal} {Phys. Rev. Lett.}\
  }\textbf {\bibinfo {volume} {111}},\ \bibinfo {pages} {106102} (\bibinfo
  {year} {2013}{\natexlab{a}})}\BibitemShut {NoStop}%
\bibitem [{\citenamefont {Ryckaert}, \citenamefont {Bellemans},\ and\
  \citenamefont {Ciccotti}(1981)}]{ryckaert1981a}%
  \BibitemOpen
  \bibfield  {author} {\bibinfo {author} {\bibfnamefont {J.-P.}\ \bibnamefont
  {Ryckaert}}, \bibinfo {author} {\bibfnamefont {A.}~\bibnamefont {Bellemans}},
  \ and\ \bibinfo {author} {\bibfnamefont {G.}~\bibnamefont {Ciccotti}},\
  }\bibfield  {title} {\enquote {\bibinfo {title} {The rotation-translation
  coupling in diatomic molecules},}\ }\href {\doibase
  10.1080/00268978100102931} {\bibfield  {journal} {\bibinfo  {journal}
  {Molecular Physics}\ }\textbf {\bibinfo {volume} {44}},\ \bibinfo {pages}
  {979--996} (\bibinfo {year} {1981})}\BibitemShut {NoStop}%
\bibitem [{\citenamefont {Ryckaert}, \citenamefont {Ciccotti},\ and\
  \citenamefont {Berendsen}(1977)}]{ryckaert1977a}%
  \BibitemOpen
  \bibfield  {author} {\bibinfo {author} {\bibfnamefont {J.-P.}\ \bibnamefont
  {Ryckaert}}, \bibinfo {author} {\bibfnamefont {G.}~\bibnamefont {Ciccotti}},
  \ and\ \bibinfo {author} {\bibfnamefont {H.~J.~C.}\ \bibnamefont
  {Berendsen}},\ }\bibfield  {title} {\enquote {\bibinfo {title} {Numerical
  integration of the cartesian equations of motion of a system with
  constraints: molecular dynamics of n-alkanes},}\ }\href@noop {} {\bibfield
  {journal} {\bibinfo  {journal} {J. Comput. Phys.}\ }\textbf {\bibinfo
  {volume} {23}},\ \bibinfo {pages} {327--341} (\bibinfo {year}
  {1977})}\BibitemShut {NoStop}%
\bibitem [{Note1()}]{Note1}%
  \BibitemOpen
  \bibinfo {note} {We recall that the $\protect \bm {R}_{\alpha }$ are fixed
  parameters in the potential. This prescription can be relaxed to include the
  location of the Gaussian charges to change but this extension of the model,
  that can be easily incorporated in the mass-zero constrained dynamics, is not
  considered here.}\BibitemShut {Stop}%
\bibitem [{Note2()}]{Note2}%
  \BibitemOpen
  \bibinfo {note} {Note that a different initialisation scheme was proposed in
  ref.~\cite {coretti2018b}. This scheme is more rigorous but its
  implementation in \protect \emph {MetalWalls}{} turned out to be impractical.
  Use of the conjugate gradient minimisation, on the other hand, is available
  in the code. This choice of initialisation did not cause visible problems and
  has negligible additional cost.}\BibitemShut {Stop}%
\bibitem [{\citenamefont {Ciccotti}\ and\ \citenamefont
  {Ryckaert}(1986)}]{ciccotti1986a}%
  \BibitemOpen
  \bibfield  {author} {\bibinfo {author} {\bibfnamefont {G.}~\bibnamefont
  {Ciccotti}}\ and\ \bibinfo {author} {\bibfnamefont {J.-P.}\ \bibnamefont
  {Ryckaert}},\ }\bibfield  {title} {\enquote {\bibinfo {title} {Molecular
  dynamics simulation of rigid molecules},}\ }\href@noop {} {\bibfield
  {journal} {\bibinfo  {journal} {Comp. Phys. Rep.}\ }\textbf {\bibinfo
  {volume} {4}},\ \bibinfo {pages} {346--392} (\bibinfo {year}
  {1986})}\BibitemShut {NoStop}%
\bibitem [{\citenamefont {Willard}\ \emph {et~al.}(2009)\citenamefont
  {Willard}, \citenamefont {Reed}, \citenamefont {Madden},\ and\ \citenamefont
  {Chandler}}]{willard2009a}%
  \BibitemOpen
  \bibfield  {author} {\bibinfo {author} {\bibfnamefont {A.~P.}\ \bibnamefont
  {Willard}}, \bibinfo {author} {\bibfnamefont {S.~K.}\ \bibnamefont {Reed}},
  \bibinfo {author} {\bibfnamefont {P.~A.}\ \bibnamefont {Madden}}, \ and\
  \bibinfo {author} {\bibfnamefont {D.}~\bibnamefont {Chandler}},\ }\bibfield
  {title} {\enquote {\bibinfo {title} {Water at an electrochemical interface -
  a simulation study},}\ }\href@noop {} {\bibfield  {journal} {\bibinfo
  {journal} {Faraday Discuss.}\ }\textbf {\bibinfo {volume} {141}},\ \bibinfo
  {pages} {423--441} (\bibinfo {year} {2009})}\BibitemShut {NoStop}%
\bibitem [{\citenamefont {Limmer}\ \emph
  {et~al.}(2013{\natexlab{b}})\citenamefont {Limmer}, \citenamefont {Willard},
  \citenamefont {Madden},\ and\ \citenamefont {Chandler}}]{limmer2013b}%
  \BibitemOpen
  \bibfield  {author} {\bibinfo {author} {\bibfnamefont {D.~T.}\ \bibnamefont
  {Limmer}}, \bibinfo {author} {\bibfnamefont {A.~P.}\ \bibnamefont {Willard}},
  \bibinfo {author} {\bibfnamefont {P.}~\bibnamefont {Madden}}, \ and\ \bibinfo
  {author} {\bibfnamefont {D.}~\bibnamefont {Chandler}},\ }\bibfield  {title}
  {\enquote {\bibinfo {title} {Hydration of metal surfaces can be dynamically
  heterogeneous and hydrophobic},}\ }\href@noop {} {\bibfield  {journal}
  {\bibinfo  {journal} {Proc. Natl. Acad. Sci. U.S.A.}\ }\textbf {\bibinfo
  {volume} {110}},\ \bibinfo {pages} {4200--4205} (\bibinfo {year}
  {2013}{\natexlab{b}})}\BibitemShut {NoStop}%
\bibitem [{\citenamefont {Willard}\ \emph {et~al.}(2013)\citenamefont
  {Willard}, \citenamefont {Limmer}, \citenamefont {Madden},\ and\
  \citenamefont {Chandler}}]{willard2013a}%
  \BibitemOpen
  \bibfield  {author} {\bibinfo {author} {\bibfnamefont {A.~P.}\ \bibnamefont
  {Willard}}, \bibinfo {author} {\bibfnamefont {D.~T.}\ \bibnamefont {Limmer}},
  \bibinfo {author} {\bibfnamefont {P.~A.}\ \bibnamefont {Madden}}, \ and\
  \bibinfo {author} {\bibfnamefont {D.}~\bibnamefont {Chandler}},\ }\bibfield
  {title} {\enquote {\bibinfo {title} {Characterizing heterogeneous dynamics at
  hydrated electrode surfaces},}\ }\href@noop {} {\bibfield  {journal}
  {\bibinfo  {journal} {J. Chem. Phys.}\ }\textbf {\bibinfo {volume} {138}},\
  \bibinfo {pages} {184702} (\bibinfo {year} {2013})}\BibitemShut {NoStop}%
\bibitem [{\citenamefont {Limmer}\ and\ \citenamefont
  {Willard}(2015)}]{limmer2015b}%
  \BibitemOpen
  \bibfield  {author} {\bibinfo {author} {\bibfnamefont {D.~T.}\ \bibnamefont
  {Limmer}}\ and\ \bibinfo {author} {\bibfnamefont {A.~P.}\ \bibnamefont
  {Willard}},\ }\bibfield  {title} {\enquote {\bibinfo {title} {Nanoscale
  heterogeneity at the aqueous electrolyte-electrode interface},}\ }\href@noop
  {} {\bibfield  {journal} {\bibinfo  {journal} {Chem. Phys. Lett.}\ }\textbf
  {\bibinfo {volume} {620}},\ \bibinfo {pages} {144--150} (\bibinfo {year}
  {2015})}\BibitemShut {NoStop}%
\bibitem [{\citenamefont {Simoncelli}\ \emph {et~al.}(2018)\citenamefont
  {Simoncelli}, \citenamefont {Ganfoud}, \citenamefont {Sene}, \citenamefont
  {Haefele}, \citenamefont {Daffos}, \citenamefont {Taberna}, \citenamefont
  {Salanne}, \citenamefont {Simon},\ and\ \citenamefont
  {Rotenberg}}]{simoncelli2018a}%
  \BibitemOpen
  \bibfield  {author} {\bibinfo {author} {\bibfnamefont {M.}~\bibnamefont
  {Simoncelli}}, \bibinfo {author} {\bibfnamefont {N.}~\bibnamefont {Ganfoud}},
  \bibinfo {author} {\bibfnamefont {A.}~\bibnamefont {Sene}}, \bibinfo {author}
  {\bibfnamefont {M.}~\bibnamefont {Haefele}}, \bibinfo {author} {\bibfnamefont
  {B.}~\bibnamefont {Daffos}}, \bibinfo {author} {\bibfnamefont {P.-L.}\
  \bibnamefont {Taberna}}, \bibinfo {author} {\bibfnamefont {M.}~\bibnamefont
  {Salanne}}, \bibinfo {author} {\bibfnamefont {P.}~\bibnamefont {Simon}}, \
  and\ \bibinfo {author} {\bibfnamefont {B.}~\bibnamefont {Rotenberg}},\
  }\bibfield  {title} {\enquote {\bibinfo {title} {Blue energy and desalination
  with nanoporous carbon electrodes: Capacitance from molecular simulations to
  continuous models},}\ }\href@noop {} {\bibfield  {journal} {\bibinfo
  {journal} {Phys. Rev. X}\ }\textbf {\bibinfo {volume} {8}},\ \bibinfo {pages}
  {021024} (\bibinfo {year} {2018})}\BibitemShut {NoStop}%
\bibitem [{\citenamefont {Haskins}\ and\ \citenamefont
  {Lawson}(2016)}]{haskins2016a}%
  \BibitemOpen
  \bibfield  {author} {\bibinfo {author} {\bibfnamefont {J.~B.}\ \bibnamefont
  {Haskins}}\ and\ \bibinfo {author} {\bibfnamefont {J.~W.}\ \bibnamefont
  {Lawson}},\ }\bibfield  {title} {\enquote {\bibinfo {title} {Evaluation of
  molecular dynamics simulation methods for ionic liquid electric double
  layers},}\ }\href@noop {} {\bibfield  {journal} {\bibinfo  {journal} {J.
  Chem. Phys.}\ }\textbf {\bibinfo {volume} {144}},\ \bibinfo {pages} {134701}
  (\bibinfo {year} {2016})}\BibitemShut {NoStop}%
\bibitem [{\citenamefont {Park}\ and\ \citenamefont
  {{McDaniel}}(2020)}]{park2020a}%
  \BibitemOpen
  \bibfield  {author} {\bibinfo {author} {\bibfnamefont {S.}~\bibnamefont
  {Park}}\ and\ \bibinfo {author} {\bibfnamefont {J.~G.}\ \bibnamefont
  {{McDaniel}}},\ }\bibfield  {title} {\enquote {\bibinfo {title} {Interference
  of electrical double layers: Confinement effects on structure, dynamics and
  screening of ionic liquids},}\ }\href@noop {} {\bibfield  {journal} {\bibinfo
   {journal} {J. Chem. Phys.}\ }\textbf {\bibinfo {volume} {152}},\ \bibinfo
  {pages} {074709} (\bibinfo {year} {2020})}\BibitemShut {NoStop}%
\bibitem [{\citenamefont {Li}\ \emph {et~al.}(2018)\citenamefont {Li},
  \citenamefont {Jeanmairet}, \citenamefont {{Mendez-Morales}}, \citenamefont
  {Rotenberg},\ and\ \citenamefont {Salanne}}]{li2018b}%
  \BibitemOpen
  \bibfield  {author} {\bibinfo {author} {\bibfnamefont {Z.}~\bibnamefont
  {Li}}, \bibinfo {author} {\bibfnamefont {G.}~\bibnamefont {Jeanmairet}},
  \bibinfo {author} {\bibfnamefont {T.}~\bibnamefont {{Mendez-Morales}}},
  \bibinfo {author} {\bibfnamefont {B.}~\bibnamefont {Rotenberg}}, \ and\
  \bibinfo {author} {\bibfnamefont {M.}~\bibnamefont {Salanne}},\ }\bibfield
  {title} {\enquote {\bibinfo {title} {Capacitive performance of water-in-salt
  electrolytes in supercapacitors: a simulation study},}\ }\href@noop {}
  {\bibfield  {journal} {\bibinfo  {journal} {J. Phys. Chem. C}\ }\textbf
  {\bibinfo {volume} {122}},\ \bibinfo {pages} {23917--23924} (\bibinfo {year}
  {2018})}\BibitemShut {NoStop}%
\bibitem [{\citenamefont {Berendsen}, \citenamefont {Grigera},\ and\
  \citenamefont {Straatsma}(1987)}]{berendsen1987a}%
  \BibitemOpen
  \bibfield  {author} {\bibinfo {author} {\bibfnamefont {H.~J.~C.}\
  \bibnamefont {Berendsen}}, \bibinfo {author} {\bibfnamefont {J.~R.}\
  \bibnamefont {Grigera}}, \ and\ \bibinfo {author} {\bibfnamefont {T.~P.}\
  \bibnamefont {Straatsma}},\ }\bibfield  {title} {\enquote {\bibinfo {title}
  {{The Missing Term in Effective Pair Potentials}},}\ }\href@noop {}
  {\bibfield  {journal} {\bibinfo  {journal} {J. Phys. Chem.}\ }\textbf
  {\bibinfo {volume} {91}},\ \bibinfo {pages} {6269--6271} (\bibinfo {year}
  {1987})}\BibitemShut {NoStop}%
\bibitem [{\citenamefont {Werder}\ \emph {et~al.}(2003)\citenamefont {Werder},
  \citenamefont {Walther}, \citenamefont {Jaffe}, \citenamefont {Halicioglu},\
  and\ \citenamefont {Koumoutsakos}}]{werder2003a}%
  \BibitemOpen
  \bibfield  {author} {\bibinfo {author} {\bibfnamefont {T.}~\bibnamefont
  {Werder}}, \bibinfo {author} {\bibfnamefont {J.~H.}\ \bibnamefont {Walther}},
  \bibinfo {author} {\bibfnamefont {R.~L.}\ \bibnamefont {Jaffe}}, \bibinfo
  {author} {\bibfnamefont {T.}~\bibnamefont {Halicioglu}}, \ and\ \bibinfo
  {author} {\bibfnamefont {P.}~\bibnamefont {Koumoutsakos}},\ }\bibfield
  {title} {\enquote {\bibinfo {title} {On the water-carbon interaction for use
  in molecular dynamics simulations of graphite and carbon nanotubes},}\
  }\href@noop {} {\bibfield  {journal} {\bibinfo  {journal} {J. Phys. Chem. B}\
  }\textbf {\bibinfo {volume} {107}},\ \bibinfo {pages} {1345--1352} (\bibinfo
  {year} {2003})}\BibitemShut {NoStop}%
\bibitem [{\citenamefont {Gingrich}\ and\ \citenamefont
  {Wilson}(2010)}]{gingrich2010a}%
  \BibitemOpen
  \bibfield  {author} {\bibinfo {author} {\bibfnamefont {T.~R.}\ \bibnamefont
  {Gingrich}}\ and\ \bibinfo {author} {\bibfnamefont {M.}~\bibnamefont
  {Wilson}},\ }\bibfield  {title} {\enquote {\bibinfo {title} {On the ewald
  summation of gaussian charges for the simulation of metallic surfaces},}\
  }\href@noop {} {\bibfield  {journal} {\bibinfo  {journal} {Chem. Phys.
  Lett.}\ }\textbf {\bibinfo {volume} {500}},\ \bibinfo {pages} {178--183}
  (\bibinfo {year} {2010})}\BibitemShut {NoStop}%
\bibitem [{\citenamefont {Wang}\ \emph {et~al.}(2014)\citenamefont {Wang},
  \citenamefont {Yang}, \citenamefont {Olmsted}, \citenamefont {Asta},\ and\
  \citenamefont {Laird}}]{wang2014}%
  \BibitemOpen
  \bibfield  {author} {\bibinfo {author} {\bibfnamefont {Z.}~\bibnamefont
  {Wang}}, \bibinfo {author} {\bibfnamefont {Y.}~\bibnamefont {Yang}}, \bibinfo
  {author} {\bibfnamefont {D.~L.}\ \bibnamefont {Olmsted}}, \bibinfo {author}
  {\bibfnamefont {M.}~\bibnamefont {Asta}}, \ and\ \bibinfo {author}
  {\bibfnamefont {B.~B.}\ \bibnamefont {Laird}},\ }\bibfield  {title} {\enquote
  {\bibinfo {title} {Evaluation of the constant potential method in simulating
  electric double-layer capacitors},}\ }\href {\doibase 10.1063/1.4899176}
  {\bibfield  {journal} {\bibinfo  {journal} {The Journal of Chemical Physics}\
  }\textbf {\bibinfo {volume} {141}},\ \bibinfo {pages} {184102} (\bibinfo
  {year} {2014})}\BibitemShut {NoStop}%
\bibitem [{\citenamefont {Jeanmairet}\ \emph {et~al.}(2019)\citenamefont
  {Jeanmairet}, \citenamefont {Rotenberg}, \citenamefont {Borgis},\ and\
  \citenamefont {Salanne}}]{jeanmairet2019b}%
  \BibitemOpen
  \bibfield  {author} {\bibinfo {author} {\bibfnamefont {G.}~\bibnamefont
  {Jeanmairet}}, \bibinfo {author} {\bibfnamefont {B.}~\bibnamefont
  {Rotenberg}}, \bibinfo {author} {\bibfnamefont {D.}~\bibnamefont {Borgis}}, \
  and\ \bibinfo {author} {\bibfnamefont {M.}~\bibnamefont {Salanne}},\
  }\bibfield  {title} {\enquote {\bibinfo {title} {Study of a water-graphene
  capacitor with molecular density functional theory},}\ }\href@noop {}
  {\bibfield  {journal} {\bibinfo  {journal} {J. Chem. Phys.}\ }\textbf
  {\bibinfo {volume} {151}},\ \bibinfo {pages} {124111} (\bibinfo {year}
  {2019})}\BibitemShut {NoStop}%
\bibitem [{\citenamefont {Carrasco}, \citenamefont {Hodgson},\ and\
  \citenamefont {Michaelides}(2012)}]{carrasco2012a}%
  \BibitemOpen
  \bibfield  {author} {\bibinfo {author} {\bibfnamefont {J.}~\bibnamefont
  {Carrasco}}, \bibinfo {author} {\bibfnamefont {A.}~\bibnamefont {Hodgson}}, \
  and\ \bibinfo {author} {\bibfnamefont {A.}~\bibnamefont {Michaelides}},\
  }\bibfield  {title} {\enquote {\bibinfo {title} {A molecular perspective of
  water at metal interfaces},}\ }\href@noop {} {\bibfield  {journal} {\bibinfo
  {journal} {Nat. Mater.}\ }\textbf {\bibinfo {volume} {11}},\ \bibinfo {pages}
  {667--674} (\bibinfo {year} {2012})}\BibitemShut {NoStop}%
\end{thebibliography}
\end{document}